# Cold Gas Plasma Sources and the Science behind their Applications in Biology and Medicine

Mounir Laroussi

Plasma Engineering & Medicine Institute, Electrical & Computer Engineering Department, Old Dominion University, Norfolk, VA 23529, USA

## Abstract

Studies on the interaction of plasma generated at atmospheric pressure and at room temperature (low temperature plasma or just cold plasma) with biological cells and tissues have revealed that cold plasma has therapeutic effects that form the basis for new medical therapies. Cold plasma exhibits bactericidal properties and at low doses can modulate cell functions, including proliferation, attachment, and migration. Research in the last two decades has shown that cold plasma can be used for wound healing and can kill cancer cells in a selective manner. This paper describes the fundamental science behind the biomedical applications of cold plasma.

## I. Introduction

Non-equilibrium plasmas are weakly ionized gases which contain charged species (electrons, positively and negatively charged ionic species), neutral species (atomic and/or molecular radicals and non-radicals), and electric fields. These plasmas also emit radiation that spans wavelengths in the infrared, visible, and ultraviolet ranges. Generally, the degree of ionization in such plasmas is less than 0.1% and their gas temperatures are relatively low, typically less than 100 °C. These plasmas can be generated at low pressures as well as at atmospheric pressure and they are often referred to as cold or low temperature plasma. Low pressure plasmas have been used for many decades in semiconductor components fabrication. Atmospheric pressure plasmas have typically been used in material surface processing applications such as making materials more hydrophilic or more hydrophobic. Since the mid-1990s atmospheric pressure low temperature plasmas have been used in biological and medical applications, which are the main focus of this paper.

Investigations on the interaction of low temperature atmospheric pressure plasmas (LTP) with biological cells and tissues showed that the effects of LTP are primarily mediated by chemically reactive oxygen species (ROS) and reactive nitrogen species (RNS) [1], [2]. Both ROS and RNS possess strong oxidative properties and can trigger signaling pathways in biological cells. Therefore, delivering controllable doses of these species to cells and tissues can lead to specific biological outcomes including the onset of apoptosis, enhancement or suppression of cell proliferation, modified cell migration, etc. The ability to locally modulate cell behavior has crucial implications in various LTP-based biomedical applications such as wound healing and cancer treatment [3] – [12]. Moreover, experiments on eukaryotic cells demonstrated that LTP can affect mammalian cells without causing thermal or other damage. For example, the ability of LTP to inactivate pathogens while keeping healthy skin cells intact is the basic concept behind the use of LTP for wound healing.

The generation and transport of the reactive species generated by LTP sources proceed in several stages. Charged species, meta-stables, and other atomic and molecular reactive species are first produced in the main plasma ignition region. These species are then transported to the discharge afterglow until finally coming in contact with the biological targets, such as cells or tissues. In the plasma afterglow, where air diffusion into the feed gas channel occurs, other secondary reactive species such as O, OH, $O_3$, $O_2^-$, NO, and $NO_2$ are generated. Figure 1 is an illustration of the processes described above [1]. Studies on the effects of plasma-generated reactive species on biological cells showed that depending on the operating conditions (power, gas type and flow rate, exposure time, distance between plasma source and target, etc.) different outcomes can be achieved. These include cell death (via necrosis or apoptosis), cell detachment, change in cellular morphology and heterotrophic pathways, change in cell motility, cell proliferation, etc. [3] – [12].

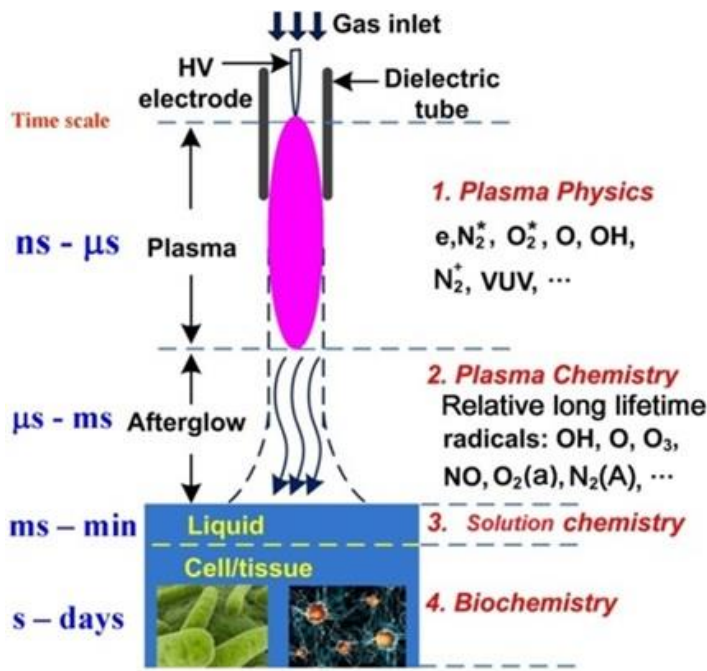

Fig. 1 Schematic depiction of low temperature plasma interacting with biological targets. Cells are typically covered by biological fluids, so the plasma first interacts with a liquid layer. Secondary and tertiary reaction by-products generated in the liquid then interact with the biological cells [1].

As shown in Figure 1 the plasma-generated reactive species first encounter a gas-liquid interface where intermediate chemical species are produced. These species then get solvated in the liquid where other reaction by-products are generated before direct interaction with the biological cells. Therefore, the rate of solvation of the various species and their lifetimes in physiological liquid need to be taken into account. Table 1 shows Henry's constant for several species of relevance, while table 2 shows the half-life of various species in physiological liquids. Note that the larger Henry's constant the deeper in the liquid a specie can penetrate with the diffusion distance roughly scaling with the square root of the half-life.

Table 1 Henry's constant for various species (source: Jet Propulsion Laboratory data base)

| Species | Henry's Constant (mol/lit x atm.) |
|---|---|
| Argon (Ar) | $1.4 \times 10^{-3}$ |
| Oxygen ($O_2$) | $1.2 \times 10^{-3}$ |
| Ozone ($O_3$) | $1.1 \times 10^{-2}$ |
| Hydroxyl (OH) | $2.9 \times 10^{1}$ |
| Hydrogen Peroxide ($H_2O_2$) | $8.4 \times 10^{4}$ |
| Nitric Oxide (NO) | $1.9 \times 10^{-3}$ |
| Nitrogen Dioxide ($NO_2$) | $4.0 \times 10^{-2}$ |

Table 2 Estimated half-life of various species in physiological environments. Note that the actual lifetime depends on factors such as pH and concentrations of reactive partners.

| Specie | Estimated half-life |
|---|---|
| OH$\bullet$ | ns |
| $NO_2\bullet$ | μs |
| $O_2^-$ | ms |
| $ONOO^-$ | ms |
| NO$\bullet$ | $1-10$ s |
| $^1O_2$ | 10 s |
| $H_2O_2$ | minutes |
| $NO_2^-$ | minutes |
| $NO_3^-$ | minutes |

The biological effects of many of the above listed reactive species are well known in cell biology. For example, the hydroxyl radical (OH) causes peroxidation of unsaturated fatty acids which are a major component of the lipids constituting the cell membrane. Another byproduct of plasma application is hydrogen peroxide ($H_2O_2$). Hydrogen peroxide possesses strong oxidative properties that affect, via the peroxide ions, lipids, proteins, and DNA. Nitric oxide (NO) is another species that can be generated by plasma as it interacts with air. The NO molecule is known to have several biological effects which include the regulation of immune-deficiencies, induction of phagocytosis, proliferation of keratinocytes, and regulation of collagen synthesis. Since plasma produces copious amounts of the above-mentioned species, it was realized early on that LTP can be used for therapeutic purposes. Many seminal experiments showed that this was indeed the case, which ultimately led to the emergence and development of the field of plasma medicine.

LTP (or cold plasma) sources that generate atmospheric pressure plasma that can safely come in contact with biological cells and tissues were developed in the 1990s and early 2000s. The biomedical applications of cold plasma emerged in the mid-1990s when LTP generated by a dielectric barrier discharge was applied to inactivate bacteria and when it was first realized that the reactive species generated by the plasma played a crucial role in the observed results [13], [14]. This groundbreaking work was followed by seminal studies on the effects of plasma on eukaryotic/mammalian cells when it was shown that low doses of LTP can cause cell detachment without causing necrosis and under some conditions apoptosis can be achieved [15], [16]. Following the early foundational works mentioned above, plasma medicine, as an emerging medical field, reached critically important milestones when the first clinical trials on wound healing and cancer treatment were conducted [7], [17].

In the last two decades the field has experienced exponential growth in term of the number of studies conducted worldwide, the number of research centers, institutes, laboratories, and groups active in the various aspects of the field as well as in terms of the large number of peer reviewed publications. Finally, and as evidence that the field has steadily been reaching a level of maturity, several books covering the fundamentals as well as the applications of LTP in medicine were published within the last decade [18] – [22].

## II. Cold Plasma Technology for Medical Therapy

### II.1 Fundamentals of Low Temperature Plasma

Nonequilibrium plasmas are weakly ionized gases where the electrons are energetic while the ions and neutrals exhibit kinetic temperatures much less than that of the electrons. Because of this particular energy distribution, the gas does not undergo heating and the discharge is said to be non-thermal. One key design feature that characterizes nonequilibrium plasma sources is the avoidance of the glow-to-arc transition. This can be achieved by various means including judicious electrodes design and/or wave-tailoring of the applied power. For example, barrier discharges use a dielectric or a resistive layer to cover one or both electrodes. This way the discharge current is self-limited, and the discharge is inhibited from turning into a spark. On the other hand, using fast rise time short pulses allows the energy to be exclusively coupled to the electron population and in this way the gas temperature remains relatively low. Pulses shorter than the glow-to-arc transition time are an ideal solution to maintain the nonequilibrium characteristic of a discharge.

Nonequilibrium, low temperature plasma offers a unique medium that can be used to interact with biotic matter to selectively induce certain biological outcomes. The effects of low temperature plasma (LTP) on biological cells are mainly mediated by its reactive oxygen species (ROS) and reactive nitrogen species (RNS) [1]. These include hydroxyl, OH, atomic oxygen, O, singlet delta oxygen, $O_2(^1\Delta)$, superoxide, $O_2^-$, hydrogen peroxide, $H_2O_2$, and nitric oxide, NO. These species can interact with cells membranes, enter the cells and increase the intracellular ROS concentrations, which may lead to DNA strands breaks, mitochondria damage, and may compromise the integrity of other organelles and macromolecules (such as proteins). ROS and RNS can also trigger cell signaling cascades, modulate cell functions, and beyond certain concentrations threshold can ultimately lead to cellular death pathways such as necrosis or apoptosis. Other LTP generated agents that may play biological roles are charged particles (electrons, negative and positive ions), photons, as well as large electric fields that may cause irreversible cellular electroporation, allowing large molecules to enter the cells.

In plasma medicine two types of low temperature plasma sources have been extensively used, the dielectric barrier discharge (DBD) and the nonequilibrium atmospheric pressure plasma jets (N-APPJ). The following are brief descriptions of these two sources.

### II.1.1 The Dielectric Barrier Discharge

One of the earliest methods to generate low temperature atmospheric pressure plasmas is to cover planar or cylindrical electrodes by a dielectric material. The electrodes are separated by a small gap where the operating gas is introduced. The electrodes are driven by voltages of several kV at frequencies in the kHz range. This electrode scheme allowed the generation of stable glow discharges that can be filamentary or diffuse depending on parameters such as frequency and gas mixture. Common dielectric materials used in DBDs include glass, quartz, ceramics and polymers. The gap distance between electrodes varies from less than 0.1 mm in plasma displays, several millimeters in ozone generators and up to several centimeters in $CO_2$ lasers. DBD devices can be made in many different geometries, including planar and cylindrical, Figure 2 shows the schematic of different types of DBD, including volumetric and surface discharges. Figure 3 shows images of a filamentary DBD and a diffuse DBD plasma.

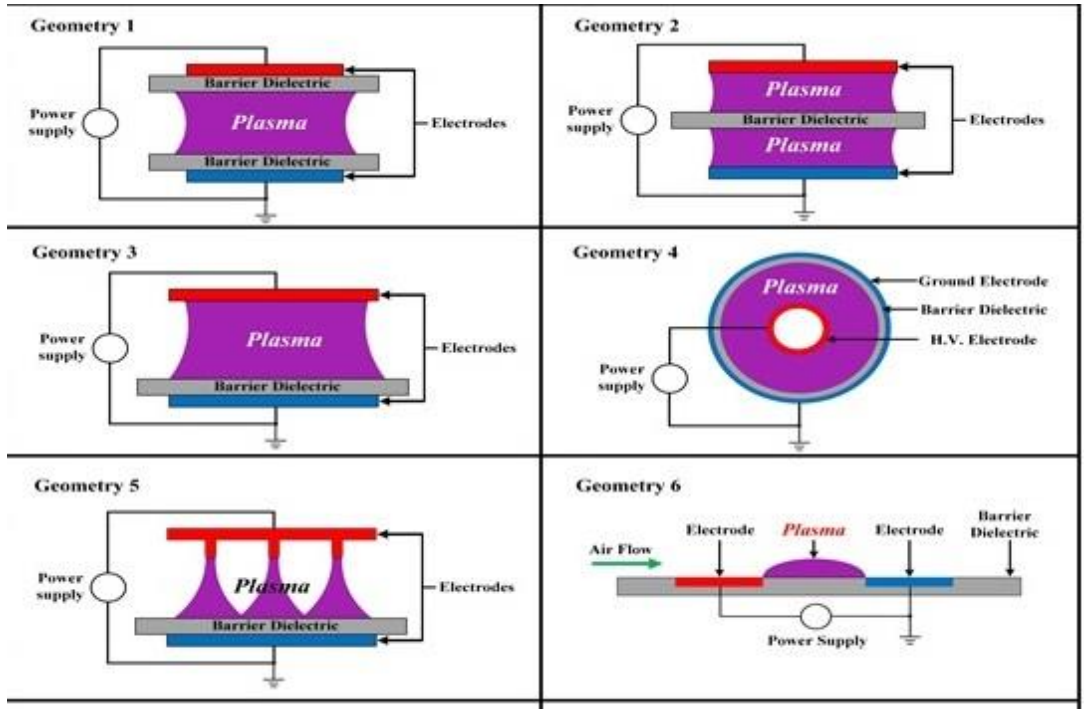

Fig.2 Dielectric barrier discharges with various configurations and electrode geometry.

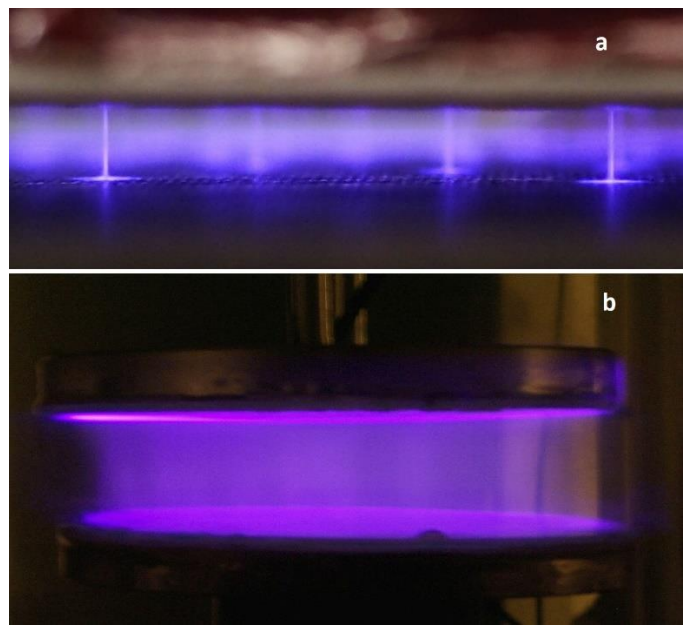

Fig. 3 Images of a DBD-generated filamentary plasma (a) and that of diffuse DBD plasma (b). Image a is by Devansh Sharma and image b is by Mounir Laroussi

Under sinusoidal excitation the electrodes of the DBD are energized by high sinusoidal voltages with amplitudes in the 1 to 20 kV range, at frequencies in the kHz range. The electrodes arrangement is generally contained within a vessel to allow for the introduction of a select gaseous mixture. When the discharge is ignited electric charges accumulate on the dielectric surface. These surface charges cause a drop in the voltage across the gas gap and a sharp decrease of the discharge current. Although generally DBDs produce filamentary plasmas, under some conditions homogeneous plasmas have been generated. The discharge current exhibits multiple current spikes per half cycle when the discharge is filamentary. However, when the discharge is diffuse the current exhibits only one current spike per half cycle. Massines et al. proposed the following mechanism which leads to the generation of a diffuse plasma. Seed electrons and metastables present between current pulses make gas breakdown possible under low electric field conditions. These seed particles can result in a Townsend-type breakdown or a glow discharge depending on the gas used. Seed electron densities greater than $10^6$ cm$^{-3}$ was found to be sufficient to keep the plasma ignited under low field conditions when helium was used [23]. It is also important to note that the surface of the dielectric can be a source of quencher species that influence the available concentration of metastable atoms. For more in-depth coverage of the DBD the reader can consult reviews [24] – [27].

The electron energy distribution function (EEDF) plays a crucial role in non-equilibrium discharges. It is through electron impact excitation and ionization that the charged particles, excited species, and radicals are produced. Therefore, by tailoring the EEDF the plasma chemistry can be controlled to a certain extent. One method to achieve such control of the EEDF was to use repetitive short high voltage pulses. These applied pulses allow for the transfer of power exclusively to the electron population. This translates in enhanced excitation and ionization. In addition, pulses with widths less than the characteristic time of the onset of the glow-to-arc transition help keep the plasma stable and inhibits it from turning into a spark. Investigators reported that for pulsed DBDs two discharges occur for every applied voltage pulse [28]. The first discharge (or "primary" discharge), was ignited at the rising edge of the applied voltage pulse while the second discharge (or "secondary" discharge), was self-ignited during the falling edge of the applied voltage pulse.

### II.1.2 Nonequilibrium Atmospheric Plasma Jets

Non-equilibrium atmospheric pressure plasma jets (N-APPJs) are devices that can emit low temperature plasma plumes in the surrounding air. These plasma sources can generate stable thin column of plasma outside the confinement of electrodes and into the surrounding environment. Because the plasma propagates away from the high voltage region and into a region where there is no externally applied electric field the plasma is electrically safe in the sense that it does not cause electrical shock to the treated biological targets. However, the plasma plume may exhibit a very high instantaneous local electric field at its tip.

N-APPJs come in various designs and geometries. Investigators have used DC, pulsed DC, RF, and microwave power to drive these plasma jets. Noble gases (such as helium and argon) and gas mixtures (such as He/air, Ar/air, He/$O_2$, Ar/$O_2$, etc.) are typically flown through at gas flow rates in the 1- 10 slm range. Figure 4 shows schematics of a few examples of N-APPJs while Figure 5 shows photographs of four different plasma jets.

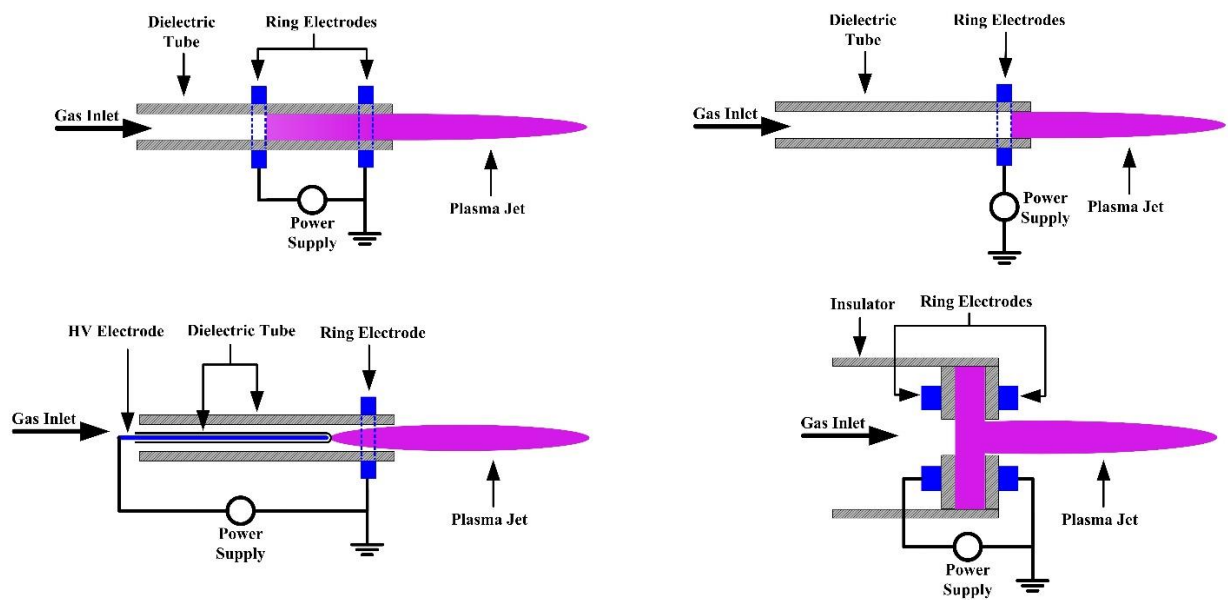

Fig. 4 Schematics of four N-APPJs with different electrodes configurations.

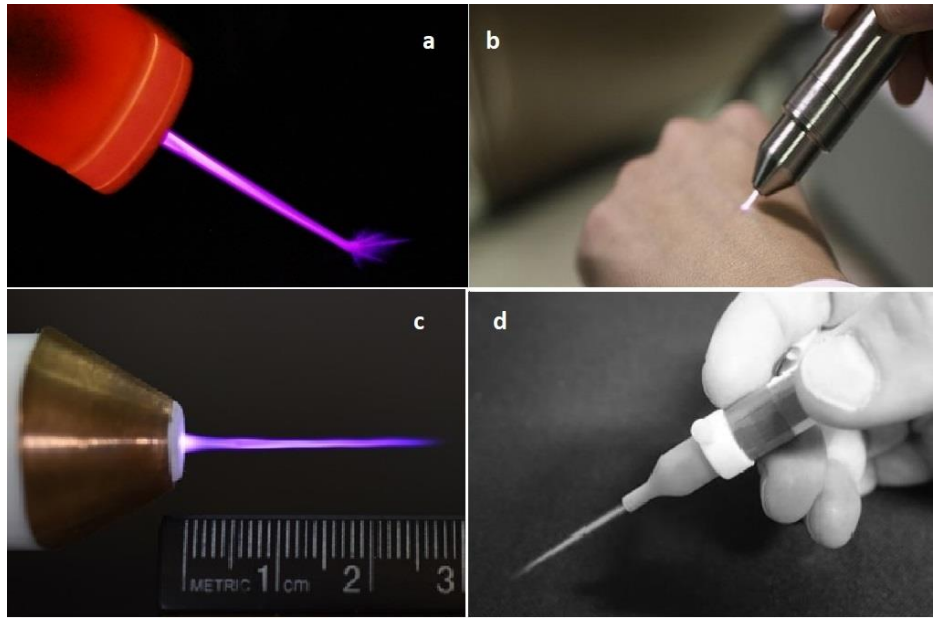

Fig. 5 Photographs of four N-APPJs. (a) the plasma pencil driven by pulsed DC; (b) the kINPen driven by RF power; (c) plasma jet driven by DC; (d) plasma jet driven by a piezoelectric transformer.

Investigators discovered that the plasma plumes emitted by N-APPJs are not continuous but are in fact discrete plasma segments that are known as "plasma bullets". These plasma bullets propagate at hypersonic velocities, up to $10^5$ m/s [29], [30]. Extensive experimental and modeling studies resulted in a good understanding of the physical mechanisms behind the generation and propagation of the plasma bullets [31]-[42]. Lu & Laroussi first proposed a photoionization model to explain the dynamics of the plasma bullets [30]. Further studies showed that the plasma bullets are in fact ionization waves which are guided by the gas channel, hence the name "guided ionization waves" [43]. The electrical field at the head of the ionization waves plays a crucial role in the propagation process and was experimentally measured by various investigators to be in the 10-30 kV/cm range [44]-[46]. Figure 6 shows a simulation of the electric field generated at the head of the plasma plume as a function of axial position and time. To learn more about N-APPJs the reader can consult reviews [47] - [50].

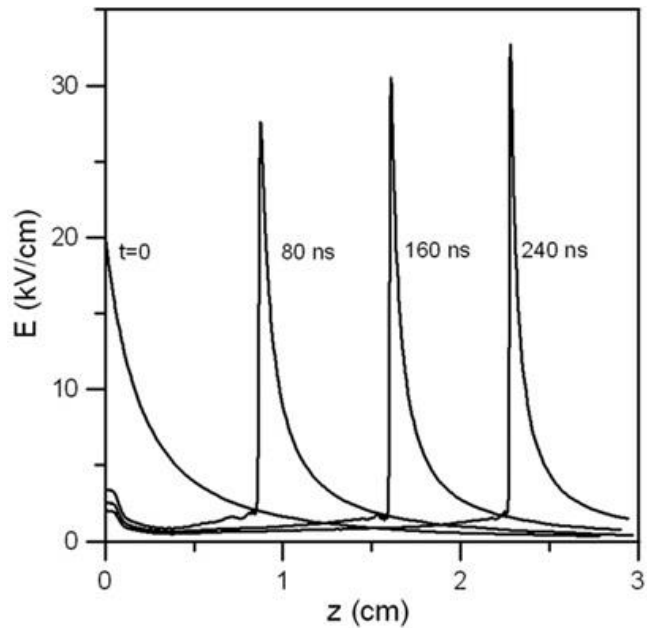

Fig. 6 Profile of the electric field at the head of the plasma plume of a pulsed plasma jet at various times and axial positions [33].

Readers who are interested in learning the basic physical laws that govern low temperature plasmas can find a thorough coverage in the foundation paper [51]. This includes discharge ignition, gas breakdown, Paschen law, diffusion, scaling laws, sheath dynamic, streamer mechanisms, etc.  In addition, the fundamental concepts behind atmospheric pressure non-equilibrium plasma, which are of great relevance to the plasma sources used in plasma medicine, can be found in the foundation paper [51]. These include scaling laws, time scales, discharge inception and breakdown mechanisms, plasma instabilities, stabilization methods, etc. [52].

### II.2 Plasma Reactive Species

DBDs and N-APPJs produce a "cocktail" of chemically reactive species in the gaseous phase including reactive oxygen species (ROS) such as hydroxyl, OH, atomic oxygen, O, singlet delta oxygen, $O_2(^1\Delta)$, superoxide, $O_2^-$, ozone, $O_3$, and hydrogen peroxide, $H_2O_2$. Reactive nitrogen species (RNS) such as nitric oxide NO, and nitrogen dioxide, $NO_2$ are also generated. When interacting with liquids other secondary and tertiary species such as nitrite, $NO_2^-$, nitrate, $NO_3^-$, peroxynitrite, $ONOO^-$, peroxynitrous acid, $HNO_3$, organic radicals, RO, are also generated.  As mentioned earlier, the concentrations and fluxes of these species play crucial roles when LTP interacts with biological matter. Therefore, determining these concentrations qualitatively and quantitatively is of paramount importance. The following provide some examples of these measurements and briefly describe the diagnostics methods used to obtain them.

As illustrative examples of LTP-produced reactive species, Figures 7 through 9 show two important species generated by plasma jets. Figure 7 shows the radial distribution of hydroxyl, OH, for different gas flow rates while Figure 8 shows the OH distribution in the presence of a target material. Figure 7 illustrates the major influence of the gas flow rate and indicates that the maximum OH density is close to the center of the plasma plume and decays with radial distance. As can be seen in Figure 8, the presence of a target influences greatly the production of OH. This highlights the fact that the type of target has a dramatic influence on the OH concentration and its radial distribution, with the maximum density at the center followed by nearly exponential radial decay. A target with dry surface leads to lower OH production while a wet surface results in a substantial increase in OH density. But it is important to keep in mind that the concentrations of species produced by plasma jets also depend on several other factors including humidity level of the surrounding environment, gas flow rate, air mole fraction, etc. Figure 9 shows the axial distribution of nitric oxide, NO, generated by an RF jet at different applied powers, NO being a very important biological molecule. Figure 9 shows that the production of NO is a function of the power applied to the plasma and that its density decays rapidly with the axial distance away from the nozzle of the device.

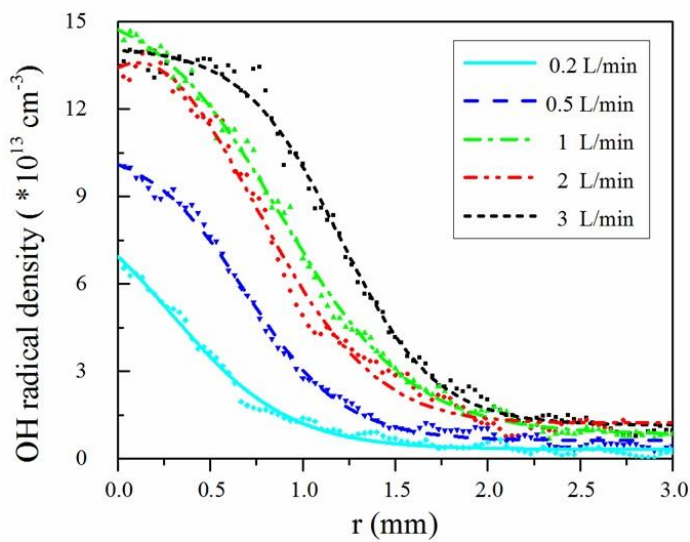

Fig. 7   Radial distribution of the OH density for various gas flow rates (gas: helium with 115 ppm of $H_2O$). The plasma jet has an insulated single pin electrode inside a quartz tube and the plasma plume is applied on top of a water film [53].

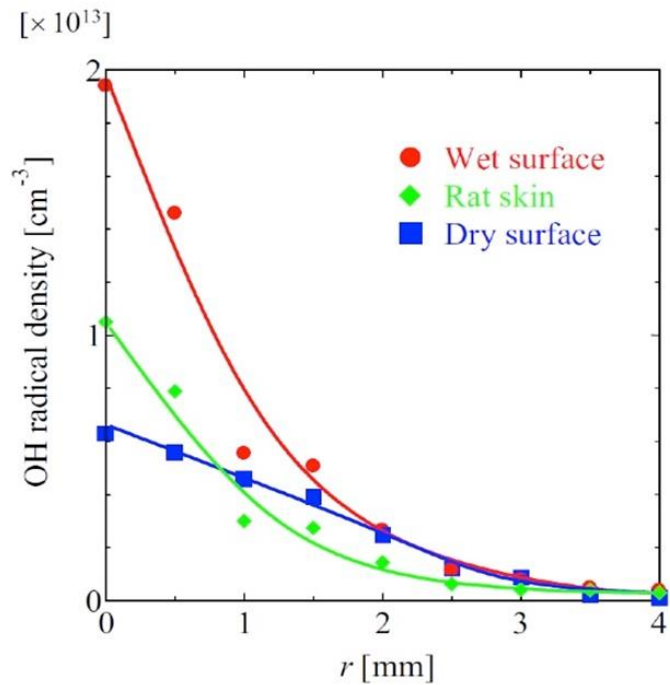

Fig. 8 OH density radial distribution on wet/dry/rat skin surfaces [54]

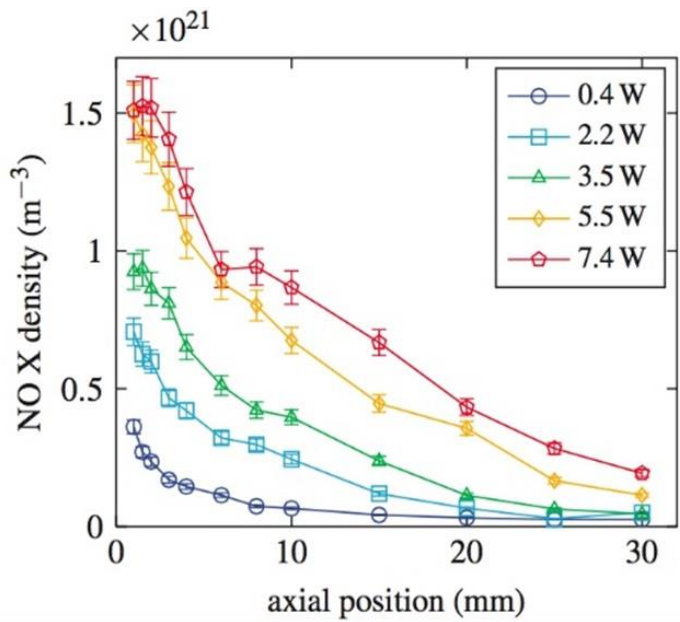

Fig. 9 Axial profiles of NO concentration for different RF discharge power. The plasma is operated at a frequency of 13.6 MHz, pulsed at 20 kHz with 20% duty cycle. Working gas is Ar at 1 L/min with 2% air admixture [55].

Finally, Figure 10 shows the concentrations of various species of biological relevance produced in the plasma plume of an argon N-APPJ source as a function of the axial position. As can be observed in the figure some important species such as O, NO, NO$_2$, O$_2^-$, O$_3$, OH, H$_2$O$_2$, and HNO$_2$ remain at relatively high concentrations in the far effluent, at up to few centimeters from the nozzle. Other species, however, are more prevalent within a short distance from the nozzle and decay to markedly low levels in the far effluent. Knowing these distributions helps inform the user which species are mostly likely to play a role and at which locations.

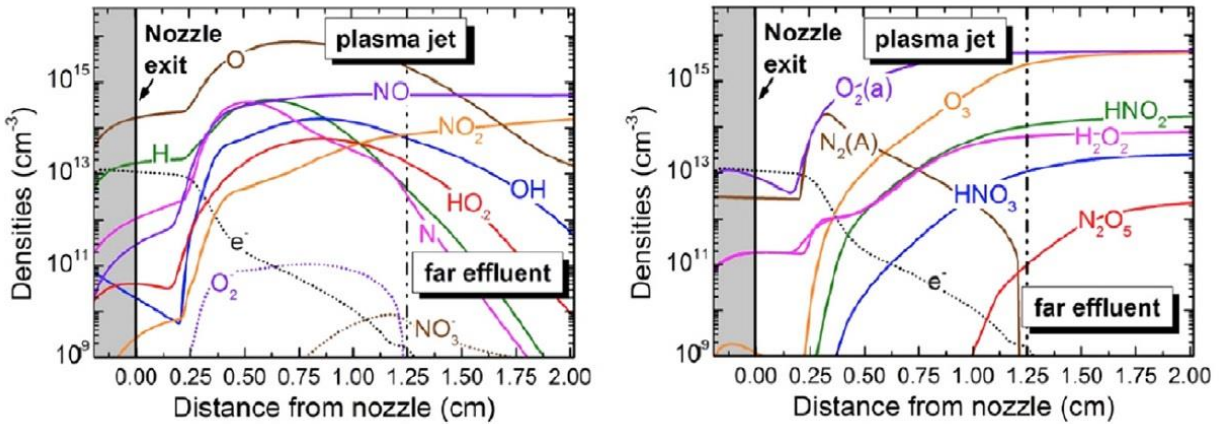

Fig, 10 Axial concentrations of biomedically active species in an Ar plasma jet [56].

To conclude this section, it is important to reiterate the fact that quantitative evaluation of the type of species, their concentrations and their axial and radial distributions is crucial for biomedical applications since the biological outcomes correlate directly with the fluxes of such species. Too high a flux of certain species could lead to distress and cell death while a low flux of the same species could lead to eustress and cell proliferation.

### III.    Diagnostics Methods for Cold Plasmas

There are several diagnostics methods to identify and measure the concentrations of reactive species produced by low temperature plasmas. Because typically LTP sources have relatively small dimensions and exhibit non-uniformities, invasive techniques (such as electrical probes) are usually not suitable. Optical diagnostics are therefore preferred since they are mostly non-invasive and can yield measurements that are resolved in time and in space. These diagnostics techniques include optical emission spectroscopy (OES), optical absorption spectroscopy, laser induced fluorescence (LIF), and two photon absorption laser induced fluorescence (TALIF). In addition, scattering techniques such as Thomson, Rayleigh, and Raman Scattering can be used. Other advanced spectroscopic methods such as cavity ring down spectroscopy have also been used to evaluate very low concentrations of some species. Finally, well-established chemical analysis techniques including mass spectrometry and electron paramagnetic resonance spectroscopy are at the investigators' disposal. In the following section only a brief coverage of OES, LIF, and TALIF diagnostic techniques is presented. For more in-depth information on optical diagnostics methods the reader can consult the review papers [57], [58].

OES is a relatively simple technique that can yield both qualitative and quantitative measurements. OES is used to identify excited species and absolute OES can measure absolute values of densities but requires careful calibration. Only excited species that emit radiation can be identified and/or measured by OES. Species in the ground state cannot be detected by OES, so they require other measurements methods such as LIF and TALIF. OES has been used to measure the gas temperature, electrons density, and electric field strength. The gas temperature is deduced from the rotational temperature in the case of molecular gases. For example, the rotational structure of the 0-0 band of the second positive system of nitrogen is often used to estimate the gas temperature. This is done by comparing experimental data with simulation data. The electron density can be found using the Stark broadening by exploring the Balmer β transition of hydrogen at 486.132 nm. However, this broadening is usually resolved only when the electron density is greater than $10^{13}$ cm$^{-3}$. OES can be used to measure the electric field by applying the polarization-dependent Stark splitting and shifting of the helium 447.1 nm line and its forbidden counterpart. To learn the details of the above mentioned OES techniques the reader can consult ref [58].

Laser induced fluorescence is useful to measure species in the ground state. LIF is based on a two-step process. First a photon is absorbed by an atom or molecule. As a result, the absorbing particle is excited to higher energy state. The excited species then undergoes spontaneous radiative decay to a lower energy level and in the process releasing a photon at a different wavelength than that of the incident laser light. The emitted photons constitute the fluorescent light signal which can be detected, and its intensity measured. Figure 11 shows an example of an LIF setup to measure the concentration of OH produced by a plasma jet. By calibration or via signal fitting process the concentration of the species can be deduced. LIF can be used to measure atomic or molecular species, however, it is more complex for molecules. This is because the radiative decay can be influenced by vibrational and rotational energy transfers. Ref [59] presents a good example of how LIF can be applied for non-equilibrium plasma.

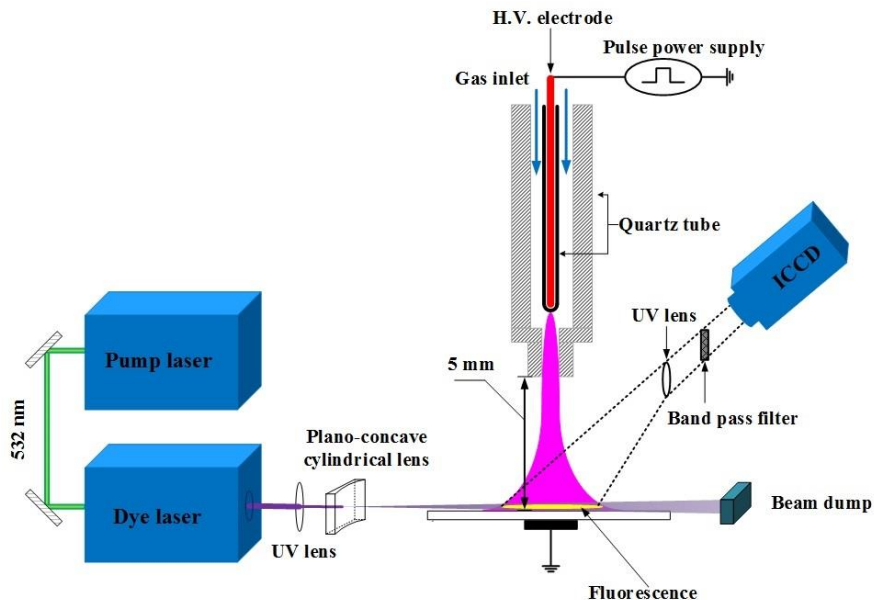

Fig. 11 LIF setup to measure the concentration of OH [53].

In two photon absorption LIF (TALIF) the probed species is first excited by a photon to an intermediate state. The intermediate level can be a virtual state or a real electronic state. A second photon excites the species into a final higher state and then after radiation decay to a lower state a fluorescence signal is emitted. As an example, TALIF has been used to measure the concentrations of atomic oxygen. In this case a laser beam at a wavelength of 225.62 nm is used to excite atomic oxygen from the $2p^4\ ^3P_2$ state to the $3p^3\ ^3P_2$ state. The fluorescence signal is captured at a wavelength of 845 nm which corresponds to the transition from $3p^3\ ^3P_2$ state to the $2s^3S$ state of O. The absolute concentration can be estimated by calibration methods. For more details the reader can consult ref [58].

The above section discussed only measurements methods to evaluate concentrations of relevant species in the gaseous phase, generated by plasma. There are well established techniques, molecular probes, and assays to carry out measurements in the liquid phase. These include spectrophotometric assays (such as Griess assay for measuring nitrate), cholorimetric/fluorometric assays (such as Amplex red assay for measuring hydrogen peroxide), singlet oxygen sensor, chemical titration (for ozone), etc.

## IV.     Treatment Modalities

In plasma medicine the biological targets (cells, tissues, organs, etc.) can be exposed to LTP in a direct or indirect way. In the case of sterilization/decontamination, the plasma can either come in direct contact with the cells (bacteria, viruses, fungi, biofilm, etc.) or the cells can be placed in a location where they are exposed only to the plasma effluent (indirect treatment). In direct treatment all the possible plasma-produced agents act on the cells. These include charged particles, reactive species, photons, electric field, and heat. In indirect treatment the contribution of charged particles, electric field, heat, and photons are greatly reduced or eliminated. In this case mostly the long-lived species generated by the plasma reach the biological target.

In the case of LTP treatment aimed for medical therapies such as wound healing or cancer treatment, direct and indirect treatments take different meanings. Direct treatment can mean that the plasma or the plasma afterglow touches the target. Indirect treatment often means that LTP is first used to activate a liquid solution and then the activated solution is applied on top of cells, wounds, or injected into tumors. The used solutions include saline solution, cell culture media, ringer's lactate solution, etc. In this treatment modality the effects of heat, photons, and electric field are eliminated. In addition, plasma activated solutions can be stored and used at a later time (hours/days). Figure 12 illustrates the steps involved in such treatment modality.

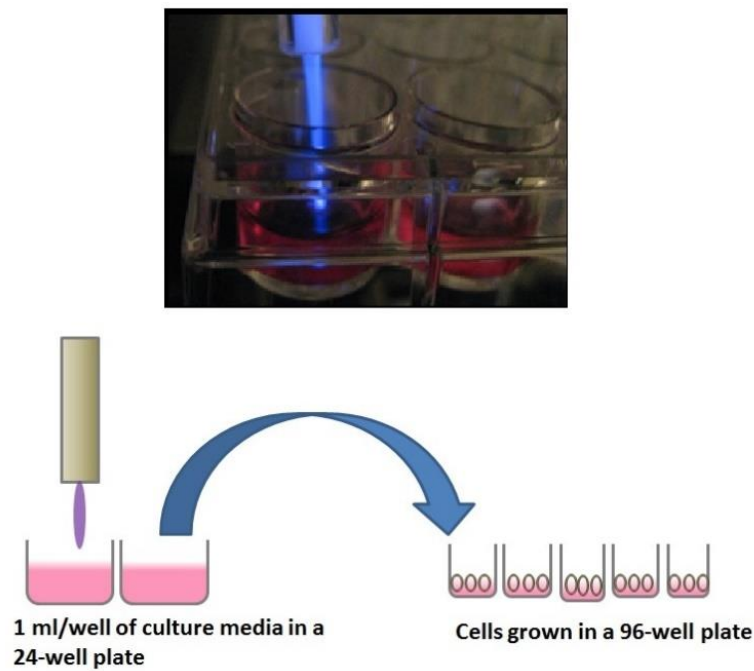

Fig. 12 Indirect treatment: First a liquid solution is exposed to LTP (top photograph), then the activated solution is applied on top of cell (bottom schematic).

### V.  Modeling of LTP-Cell and LTP-Tissue Interactions

This section briefly describes the computer models that have been used to simulate the interaction of LTP with biological matter. For extensive descriptions of the computer simulations used to model non-equilibrium low temperature plasmas the reader is referred the foundation paper by Alves et al. [60].

A few investigators have attempted to model the interactions of low temperature plasma with biological cells and tissues. So far, two types of simulations approaches have been used, molecular dynamics (MD) simulations and hydrodynamic simulations. MD simulation (both classical MD and *ab initio* MD) is a method that allows the system to be described on the atomic level while hydrodynamic simulation is used on a more macroscopic level to describe the system on the cellular/tissue level. In the latter simulation method, the cells and tissues are modeled by dielectric materials having certain permittivities and conductivities. Using the above-mentioned methods, simulations of plasma interactions with bacteria, with eukaryotic/mammalian cells, and with tissues have been carried out [61], [62]. For example, MD simulations have been used to study the plasma effects on DNA, peptidoglycan, lipid bilayer, ion transport, and electroporation. These simulation studies are very important as they can shed light on phenomena on the atomic, molecular, sub-cellular, and cellular levels and can yield useful data that are hard or even impossible to obtain experimentally. The following are a few examples of what has been uncovered using such computer simulations.

To elucidate the effects of LTP on bacteria on fundamental and molecular levels, Yusupov et al. conducted computer simulations on the interaction of ROS with bacterial peptidoglycan, which is a major constituent of the bacteria cell wall [61]. For this they used reactive molecular dynamics simulations and they found that ROS affect the structure of peptidoglycan by breaking the C-N, C-O, and C-C bonds. Table 3 shows the percentage of breaking events of the C-N, C-O, and C-C bonds and which plasma species is responsible for the breakage [61]. Note that, for example, hydrogen peroxide, $H_2O_2$, was found to be responsible for breaking the C-O and C-C bonds, which indicate that $H_2O_2$ specifically affects the disaccharide part of the peptidoglycan. For detailed description of the computational method used and all the simulation results the reader is referred to ref [61].

**Table 3. Fraction of Important Bond Dissociations (i.e., C−N, C−O, and C−C Bonds) and Associated Standard Deviations upon Impact of O, O$_3$, OH, and H$_2$O$_2$$^a$, [61]**

| incident plasma species | C−N bond breaking events (%) | ether C−O bond breaking events (%) | C−C bond breaking events (%) |
|---|---|---|---|
| O atoms | 26 ± 6 | 78 ± 6 | 38 ± 7 |
| O$_3$ molecules | 8 ± 4 | 56 ± 7 | 26 ± 6 |
| OH radicals | 8 ± 4 | 54 ± 7 | 14 ± 5 |
| H$_2$O$_2$ molecules | 0 | 44 ± 7 | 12 ± 5 |

$^a$Note that the values are calculated from 50 independent simulations for each incident species.

Chen et al. developed a molecular simulation of plasma chemistry that can link to functions on the cell and tissue levels [63]. In the case of biofilms, a reactive penetration model was used for mass transfer of transient plasma reactive species across the gas-liquid boundary. In the case of tissue, they used a fluid model and equations of heat and electric field transfer through the skin. Their results revealed that the penetration of plasma chemistry into hydrated biofilms is 10 -20 μm deep, which correlated with penetration of liquid-phase plasma chemistry dominated by ROS. With their model they were also able to assess the thermal and electrical safety of LTP devices used on living tissue [63].

MD simulations have been used to study electro-permeabilization of cell membranes, ion transport through electropores, reactive species interaction with cell membranes and with biomolecular structures, lipid peroxidation, and interaction of reactive species with liquids. Detailed coverage of the computational methods and the results obtained in relations to these interactions can be found in ref [62].

Babaeva and Kushner carried out hydrodynamic simulations to study the propagation of electric field, plasma filaments through tissue, and fluxes of reactive species on tissues [64] – [66]. They used a 2D plasma hydrodynamic nonPDPSIM to study the propagation of streamers, the production of reactive species and charged species and their fluxes on tissues and wounds. Using this model they also investigated electroporation and the propagation of electric fields through skin and wounds. They found that the curvature of the skin influences greatly the propagation of the plasma filaments. As the filaments reach the skin surface charge accumulation occurs resulting in the production of a lateral electric field which spreads the filaments over the surface of the wound. In addition, significant electric field penetration into the intracellular structure can occur to establish a high enough field for electroporation to occur. These investigators also quantified the fluxes of ROS and RNS, ions and photons towards the wound. They found that for typical DBD operating conditions, both the ROS and RNS show fluences over a treatment time of a few seconds comparable to those on the surface [64] – [66]. The same investigators also reported on the interaction of ROS and RNS with liquid-covered wounds and showed that alkane-like hydrocarbons in the liquid influences which species reach the wound.

## VI.    Conclusion

The advent of the development of atmospheric pressure low temperature plasma sources has enabled the introduction of plasma technology to the life sciences. Here plasma is used to overcome several hygiene and medical challenges, which include the inactivation of antibiotic resistant pathogens, the healing of chronic wounds, and the destruction of various cancer cell lines both *in vitro* and *in vivo*. The biological effects of plasma are mostly mediated by the reactive oxygen and nitrogen species it produces. Therefore, judicious control of the production rates of these species is crucial to achieving reliable and reproducible results. The recent clinical trials for wound healing and cancer treatment show that cold plasma technology can indeed play a role in various medical therapies. One great advantage of using cold plasma in medicine is the lack of serious side effects to the patient.